\documentclass[]{spie}  
\usepackage[]{graphicx}
\usepackage{caption}
\usepackage{subcaption}

\title{Large Binocular Telescope Interferometer Adaptive Optics: On-sky performance and lessons learned} 

\author{
Vanessa P.\ Bailey\supit{a}, 
Philip M.\ Hinz\supit{a}, 
Alfio T.\ Puglisi\supit{b},
Simone Esposito\supit{b},
Vidhya Vaitheeswaran\supit{a},
Andrew J. Skemer\supit{a},
Denis Defrere\supit{a},
Amali Vaz\supit{a},
Jarron M.\ Leisenring\supit{a}
\skiplinehalf
\supit{a}Steward Observatory, University of Arizona, 933 N Cherry Ave., Tucson, AZ, USA; \\
\supit{b}Osservatorio Astrofisico di Arcetri, Largo E. Fermi 5, 50125 Firenze, Italy
}

\authorinfo{Send email correspondence to VB at vbailey@as.arizona.edu}

\begin{document} 
  \maketitle 

\begin{abstract} 
The Large Binocular Telescope Interferometer is a high contrast imager and interferometer that sits at the combined bent Gregorian focus of the LBT's dual 8.4~m apertures. The interferometric science drivers dictate 0.1'' resolution with $10^3-10^4$ contrast at $10~\mu m$, while the $4~\mu m$ imaging science drivers require even greater contrasts, but at scales $>$0.2''. In imaging mode, LBTI's Adaptive Optics system is already delivering $4~\mu m$ contrast of $10^4-10^5$ at $0.3''-0.75''$ in good conditions. Even in poor seeing, it can deliver up to 90\% Strehl Ratio at this wavelength. However, the performance could be further improved by mitigating Non-Common Path Aberrations. Any NCPA remedy must be feasible using only the current hardware: the science camera, the wavefront sensor, and the adaptive secondary mirror. In preliminary testing, we have implemented an ``eye doctor'' grid search approach for astigmatism and trefoil, achieving 5\% improvement in Strehl Ratio at $4~\mu m$, with future plans to test at shorter wavelengths and with more modes. We find evidence of NCPA variability on short timescales and discuss possible upgrades to ameliorate time-variable effects.
\end{abstract}



\section{Scientific Motivation}
\label{sec:intro}  

Exoplanetary science is maturing as a field, to the point where not only detection but characterization of exoplanets and their environments is possible. While transit and radial velocity surveys provide population statistics for inner planets, and the opportunity to characterize the most closely orbiting ones, they cannot yet characterize in detail planets located in the habitable zones of extrasolar systems, much less those in the outer reaches. Direct imaging and spectroscopy are already utilized to characterize the properties of gas giant planets in wide orbits, and are likely to play a significant role in the future characterization of terrestrial analogues.

High-contrast, high-resolution instrumentation is necessary to achieve these goals. In the outer exosolar systems, imaging gas giant exoplanets around their bright host stars requires contrasts of $10^4 - 10^6$ at separations of 0.75'' or less.  Future terrestrial planet finding missions searching for reflected visible light from Earth-like planets will require contrasts of $10^9$ or more at even smaller inner working angles. At this level, the aggregate light reflected by the dust co-orbiting with the planet, it's ``exozodiacal light,'' can swamp the planet's signal. It is essential to understand the \textit{exozodi} population of nearby stars in order to design effective terrestrial planet finding surveys in the future.

The Large Binocular Telescope Interferometer (LBTI \cite{Hinz2008, Hinz2008a, Hinz2012, Hinz2014}) will be used to characterize both the exozodi and giant planet populations around nearby stars.  The Hunt for Observable Signatures of Terrestrial Systems (HOSTS) survey \cite{Danchi2014} will measure the $10~\mu m$ thermal emission from exozodiacal dust. To derive meaningful population statistics, LBTI much reach sensitivities of 10--30 \textit{zodis} (1 zodi is defined as the integrated flux of Earth's zodiacal light); this requires $10~\mu m$ contrasts of up to $10^4$ at 0.1''.  Simultaneously, in order to set detection limits of $1-5$ Jupiter masses at $<100$~AU, the LBTI Exozodi Exoplanet Common Hunt (LEECH) survey\cite{Skemer2014} will require $4~\mu m$ contrasts of $10^3 - 10^5$ at separations of $0.2 - 1''$.

These two science drivers dictate the design of LBTI and its Adaptive Optics (AO) system. LBTI sits at the combined bent Gregorian focus of the Large Binocular Telescope's (LBT) two 8.4~m mirrors. It consists of two science channels: a $1-5~\mu m$ camera (LMIRCam \cite{Skrutskie2010, Leisenring2012}) and an $8-13~\mu m$ camera (NOMIC \cite{Hoffmann2014}). Exoplanet detection can be achieved with ``conventional'' high-contrast direct imaging at $1-5~\mu m$. In this mode, the light from the two apertures is not coherently combined, but instead independently imaged on the LMIRCam focal plane. The two images may be either overlapped or separated, depending on whether raw sensitivity or redundant PSF measurement is more important to the particular observing goal. However, the resolution set at $10~\mu m$ by the 8.4~m apertures is too coarse to probe typical exozodi spatial scales. Therefore, LBTI coherently combines the light from the two LBT apertures to create an interferometer with a baseline of 23~m for this application\cite{Defrere2014}. Both applications require exquisite wavefront control and low thermal background contribution from the optical elements; these requirements shape the design of the AO system.

In this paper we describe the LBTI AO system hardware and current performance. We detail the effects of Non-Common Path Aberrations (NCPA) on image quality and our current mitigation strategy. Finally, we describe limitations of our current approach and our future plans for NCPA mitigation.


\section{LBTI AO Hardware} 

LBTI AO consists of two independent natural guide star adaptive optics systems, one for each LBT aperture. These are near-clones of the LBT First Light AO (FLAO) systems \cite{Esposito2011}. Each system consists of a dedicated visible light pyramid wavefront sensor (PWFS) and an Adaptive Secondary Mirror (ASM). A dichroic entrance window to LBTI sends the $<1~\mu m$ wavelength light directly into the PWFS while transmitting the IR light to LBTI. In dual-aperture ``binocular'' observing mode, each AO system is operated independently. The telescope control software executes the commands necessary for binocular pointing.

Rather than opting for the traditional AO layout, where deformable mirrors are added to the existing optical train, the LBT has incorporated its deformable mirrors directly into the telescope design in the form of ASMs \cite{Riccardi2010, Christou2014}.  The ASMs can be set in either adaptive mode or fixed mode and are therefore used by both AO and seeing-limited LBT instruments. In other words ASMs nominally replace fixed secondary mirrors at the LBT for all applications. The primary benefit for LBTI of this scheme is that it adds no additional warm optics in the system \cite{Lloyd-Hart2000}. With only three warm optics ahead of LBTI (the primary, secondary and tertiary mirrors) the telescope contribution to the thermal IR background is minimized.  

The ASM itself is a voice coil-actuated thin shell design that is robust against actuator failure and can be calibrated on the telescope.  Each 0.9~m diameter mirror has 672 voice coil actuators that float the 1.6~mm Zerodur shell above its reference body. Voice coil actuators have a large stroke, negating the need for a ``woofer/tweeter'' deformable mirror pair, although during closed loop operation, tip/tilt and low order aberrations are periodically offloaded to the telescope active optics to free up ASM stroke. The voice coil design is robust against actuator failure. In contrast to microelectromechanical (MEM) mirrors, where failed actuators ``stick'' in place, failed voice coil actuators ``float,'' meaning that although they cannot be actively controlled, they smoothly follow the local mirror figure.  Another advantage to the LBT ASM design is that the AO system can be end-to-end calibrated in situ. This is because the LBT is a Gregorian design, so by placing a retroretlecting optic  at the ASM focus, the ASM can be illuminated by an artificial source built into the wavefront sensor. Our reconstructors are built with a Karhunen Loeve modal basis\cite{Riccardi2010}, and we have successfully controlled up to 400 modes on sky.

\begin{figure}[tb]
\centering
\begin{subfigure}{.5\textwidth}
  \centering
  \includegraphics[width=\textwidth]{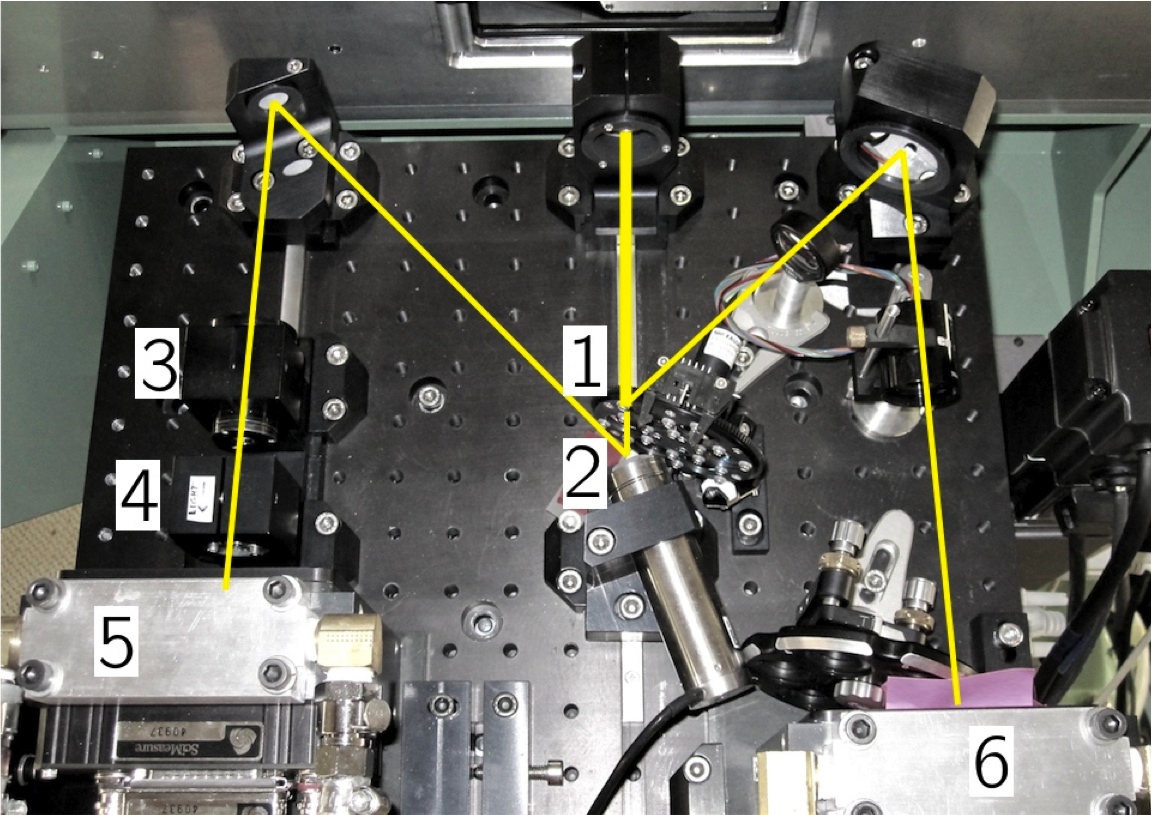}
\end{subfigure}
\begin{subfigure}{.4\textwidth}
  \centering
  \includegraphics[width=.9\textwidth]{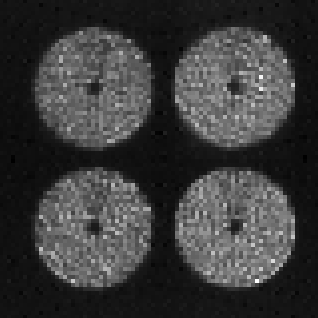}
\end{subfigure}
\caption{
\textbf{Left} LBTI pyramid wavefront sensor unit with key components labeled. Yellow lines indicate the light path. After entering the input triplet lens, the beam is split at (1), with several beamsplitter options available depending on the guide star brightness. On the right arm, the beam is imaged onto a ``viewfinder'' CCD (6) with a field of view of $\sim20''$. On the left arm, the beam comes to a focus at the tip of the pyramid optic (3) and is then re-imaged by lens (4) to form pupil images on the wavefront sensor CCD (5). The piezo actuated mirror (2) modulates the beam angle so that the PSF circumscribes the tip of the pyramid once per CCD frame; this increases the dynamic range of the WFS by preventing slope saturation.  
\textbf{Right} Example on-sky WFS pupil images taken with 400 modes of correction on a bright star. The dark wedge at 12 o'clock is the shadow of the support arms holding the secondary and tertiary mirrors.}
\label{fig:WUnit}
\end{figure}

LBTI, like any AO instrument on LBT, has a dedicated wavefront sensor. Figure \ref{fig:WUnit} details the optical layout. Our visible light PWFS is nearly identical to the FLAO PWFS \cite{Tozzi2008a, Esposito2010}, except that because LBTI has a fixed orientation relative to the telescope, we do not need pupil rerotating optics. The heart of the unit is a four-sided pyramid-shaped optic placed at a focal plane in the WFS, where it splits the PSF into 4 quadrants (implementing the Foucault knife-edge test in X and Y simultaneously). These four quadrants are reimaged into four pupil images on the CCD. The relative intensities of the pixels at corresponding locations in the four pupil images is determined by the local slope of the wavefront.  A modulating tip/tilt mirror placed before the pyramid modulates the PSF around the tip of the pyramid, with typical amplitudes of several $\lambda/D$, to increase the dynamic range of the system by preventing slope saturation.

The most distinct advantage of PWFSs is that the number of subapertures is dictated solely by the number of pixels across the pupil images; variable CCD binning makes the number of subapertures a dynamic quantity. With this flexibility, LBTI can guide on the brightest stars in the sky as well as those with $R\sim16$~mag. CCD read noise sets the faint limit. Of course, with fewer subapertures, fewer spatial modes can be sensed and therefore corrected, so overall performance is lower on fainter stars. On sky LBTI AO can correct 400, 153, 66, and 36 modes on stars with R magnitudes of approximately $<9$, $<11.5$, $<13.5$, and $<16$, respectively. On guide stars at the bright end of each binning range, the system can run at up to 990~Hz, with decreasing frequency on guide stars at the faint end of each binning range.


\section{Performance} 

\begin{figure}
\centering
\begin{subfigure}{.5\textwidth}
  \centering
  \includegraphics[width=\textwidth]{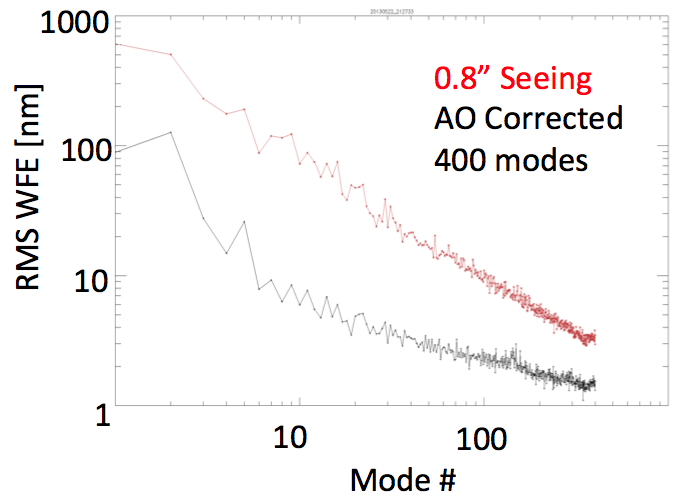}
\end{subfigure}%
\begin{subfigure}{.5\textwidth}
  \centering
  \includegraphics[width=.85\textwidth]{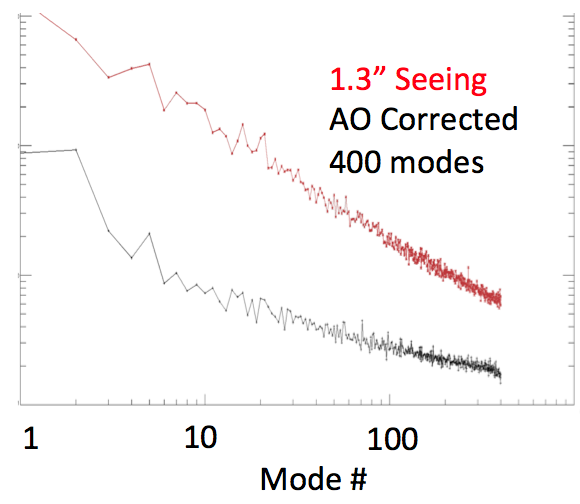}
\end{subfigure}
\caption{RMS Wavefront Error vs.\ mode number for two different seeing conditions, as measured by the wavefront sensor. In red is the WFE of the native seeing, and in black is that of the AO-corrected wavefront. The AO system achieved good rejection at all modes, with RMS amplitudes of 1-2~nm at the highest spatial frequencies. }
\label{fig:WFE}
\end{figure}

On bright guide stars, LBTI AO achieves good wavefront error (WFE) rejection on all corrected spatial scales, reaching RMS WFE as low as 1~nm at mode 400. Figure \ref{fig:WFE} shows examples of WFE as measured by the wavefront sensor for 0.8'' seeing and 1.3'' seeing. From seeing-limited to AO-corrected, the integrated WFE in these plots decreases by a factor of $\sim5$ and 9 in the two cases. Tip/tilt correction is not as robust as that of other low order modes, likely due to high-frequency vibration.  The datasets plotted average together the WFS readings over $\sim1$~sec, and so some of this tip/tilt jitter is frozen out in the short science frames. 

It is important to note that the optical gain of a modulating pyramid varies depending on the modulation amplitude, and that variable PSF quality is qualitatively equivalent to variable modulation. Therefore, as the quality of correction changes due to natural fluctuations in seeing, the optical gain of the system changes. We do not yet measure these optical gain variations (see \ref{sec:Future}). The the AO corrected WFE plotted in Figure \ref{fig:WFE} is calculated assuming an optical gain equivalent to when the reconstructor was generated. This is a good approximation for high Strehl Ratio data like these, but becomes an increasingly poor approximation for poor seeing or faint-star data.

\begin{figure}
\centering
\includegraphics[width=.7\textwidth]{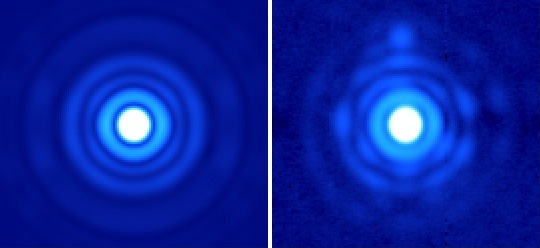}

\caption{\textbf{Left} Theoretical $L$-band PSF. \textbf{Right} $L$-band image of a bright star, after ``eye doctor'' NCPA correction (discussed in Sect. \ref{sec:eyeDoctor}). During the 2~min integration, the seeing varied between 1.4'' and 1.9''. The AO system was running with 300 spatial modes at 990~Hz, and the final image had a Strehl Ratio of 89\%. The source at outside the third Airy ring 12 o'clock is an optical ghost.}
\label{fig:PSF}
\end{figure}

The resulting $4~\mu m$ images on the short wavelength science camera, LMIRCam, have good image quality and stability over a range of seeing conditions, routinely achieving Strehl Ratios (SRs) of 85\% or higher.  We measure SR by first creating a theoretical broadband PSF from the observed LBT aperture, including the effects of the central obscuration and secondary and tertiary mirror supports. We shift the theoretical PSF to align with the observed PSF to within 0.1~px. Depending on the signal to noise (S/N) of the science image, we measure the integrated flux in apertures ranging from the 1st to 5th Airy ring. Based on tests of high S/N, high SR images, the uncertainty in SR introduced by measuring integrated flux with a small aperture is 1-3\%. Figure \ref{fig:PSF} shows the theoretical LMIRCam PSF in the broadband L filter ($3.4-4~\mu m$) and an LMIRCam image of a bright star created from 2~min of stacked 0.3~sec images. The images were not shifted before coadding in order to preserve the effects of tip/tilt jitter. During this dataset the AO system was correcting 300 modes at 990~Hz. Although the seeing was varying between 1.4'' and 1.9'', the SR was 89\%. 

\begin{figure}
\centering
\includegraphics[width=.8\textwidth]{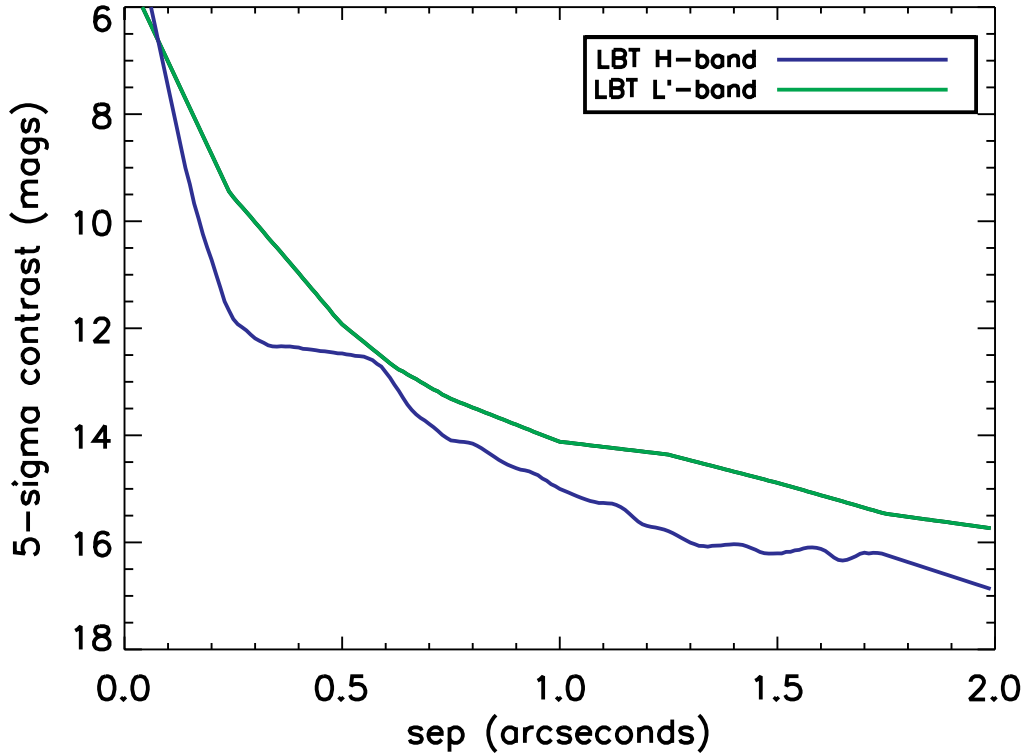}
\caption{Contrast curve for bright stars observed with LMIRCam/LBTIAO (green) and PISCES/FLAO (blue). Datasets are comparable in terms of rotation ($\sim60^\circ$) and wall clock time ($\sim2$~h). LBTI is delivering the highest contrast $L$-band images of any instrument to date.}
\label{fig:contrast}
\end{figure}

For longer Angular Differential Imaging datasets, typical of LBTI direct imaging observing sequences, we achieve unprecedented $L$-band contrasts, competitive with $H$-band results. Figure \ref{fig:contrast} shows a representative contrast curve from data obtained as part of the LEECH survey. In green is an $L$-band contrast curve from LMIRCam using the LBTI AO system\cite{Skemer2014}. In blue is an $H$-band contrast curve from PISCES observations using the (nearly identical) FLAO system\cite{Skemer2012}. The datasets had comparable sky rotation and integration: $\sim2$~hr of wall clock time and $\sim60^\circ$ of rotation. The effect of the control radius ``knee'' is prominent in the PISCES data, but more muted in the $L$-band data, due to a combination of higher sky background noise contribution and the more sparsely sampled contrast curve at L-band.  

\subsection{NCPA} 
\label{sec:ncpIntro}

\begin{figure}
\centering
\begin{subfigure}{.3\textwidth}
  \centering
  \includegraphics[width=.8\textwidth]{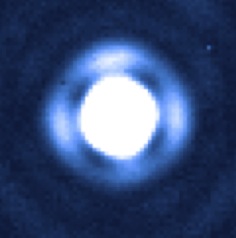}
\end{subfigure}%
\begin{subfigure}{.3\textwidth}
  \centering
  \includegraphics[width=.8\textwidth]{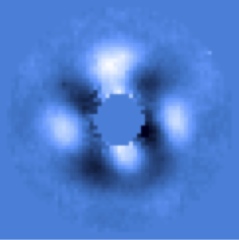}
\end{subfigure}
\begin{subfigure}{.3\textwidth}
  \centering
  \includegraphics[width=.8\textwidth]{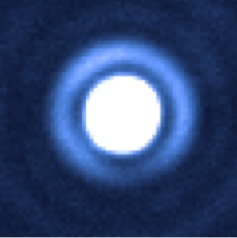}
\end{subfigure}
\caption{Effects of Non-Common Path Aberrations on LMIRCam $4~\mu m$ PSFs. \textbf{Left:} Original PSF with SR$\sim85$\%. \textbf{Middle:} Residuals after subtraction of azimuthally averaged PSF profile, highlighting asymmetries in the first Airy ring.  \textbf{Right:} PSF after using ``eye doctor'' routine to correct only astigmatism and trefoil ($\sim200$~nm RMS total), yielding SR$\sim90$\%.}
\label{fig:NCPA}
\end{figure}

For bright guide stars and average-to-good seeing conditions, the primary factor limiting LMIRCam image quality is not the AO-corrected wavefront but Non-Common Path Aberrations (NCPA). The left panel of Figure \ref{fig:NCPA} shows an example LMIRCam $4~\mu m$ PSF, and the right panel shows the residuals after subtraction of the PSF azimuthal average. The prominent cross pattern is the result of $\sim200$~nm RMS of astigmatism; the effects of $\sim50$~nm of trefoil are less pronounced. The overall Strehl Ratio of the PSF is $\sim85$\%; typical $4~\mu m$ SRs range from 80-90\%. Theoretical SRs at $4~\mu m$ exceed 95\%, so mitigating NCPA could significantly impact the image quality delivered. Improving the SR would not only increase the core flux (increase the peak signal on faint companions), but also decrease the flux in the first Airy ring and beyond (decrease the photon noise and variability at small inner working angles). In order to remain competitive with other extreme-AO systems such as GPI\cite{Macintosh2008}, SPHERE\cite{Dohlen2006}, SCExAO\cite{Guyon2010}, and P1640\cite{Hinkley2011}, LBTI must implement NCPA correction.

\section{Mitigating Non-Common Path Aberrations} 

LBTI does not have dedicated NCPA measurement hardware. We explore solutions that can be executed using the science camera and WFS alone to sense NCPA.

\subsection{Eye Doctor} 
\label{sec:eyeDoctor}

\begin{figure}
\centering
\includegraphics[width=.8\textwidth]{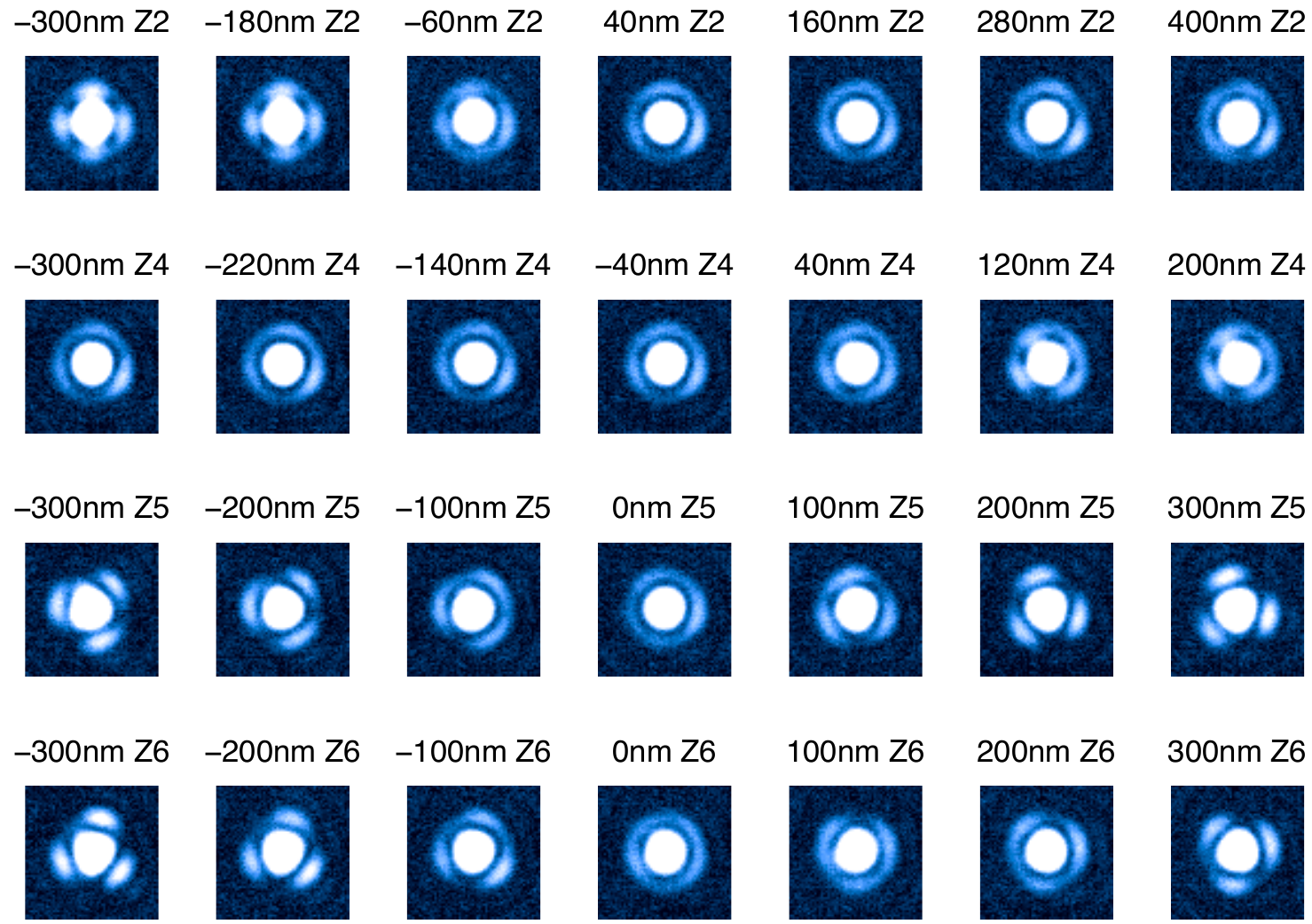}
\caption{Example LMIRCam images (log stretch) from an ``eye doctor'' grid search of astigmatism and trefoil. The title above each postage stamp is the RMS amplitude and mode number. The varying asymmetry of the first Airy ring is the most visible signature of the injected aberrations.}
\label{fig:eyedoctorrun}
\end{figure}

We have implemented a simple method for measuring NCPA that we dub the ``eye doctor'' approach. We run this routine during set-up for observations. We execute a grid search over one or more low-order aberrations to determine the amplitude that yields the best PSF on the science camera. We first inject aberrations into the AO system by modifying the WFS reference slope offsets, then take a short ($<1$~sec) science camera exposure of the resulting PSF. We build slope offsets from the system AO reconstructor itself; the low order Karhunen Loeve modes approximate astigmatism, trefoil, etc.  We search one mode at a time, applying the best amplitude of a given mode before beginning the search on the next mode. Figure \ref{fig:eyedoctorrun} shows an example run including astigmatism and trefoil. Note that a larger amplitude of astigmatism than trefoil is necessary to appreciably impact the PSF morphology.  

\begin{figure}
\centering
\begin{subfigure}{\textwidth}
  \centering
  \includegraphics[width=.7\textwidth]{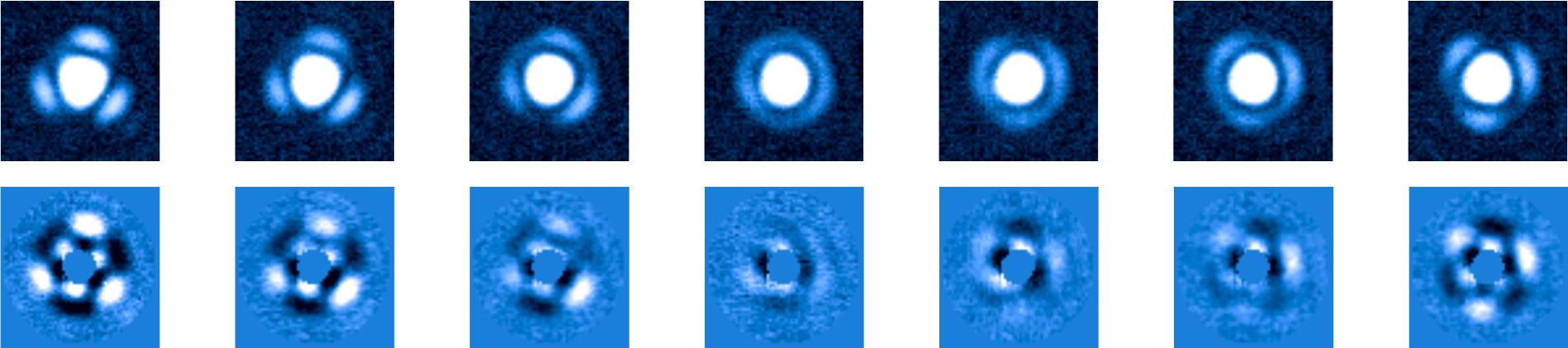}
\end{subfigure}
\newline  \newline  \newline
\begin{subfigure}{\textwidth}
  \centering
  \includegraphics[width=\textwidth]{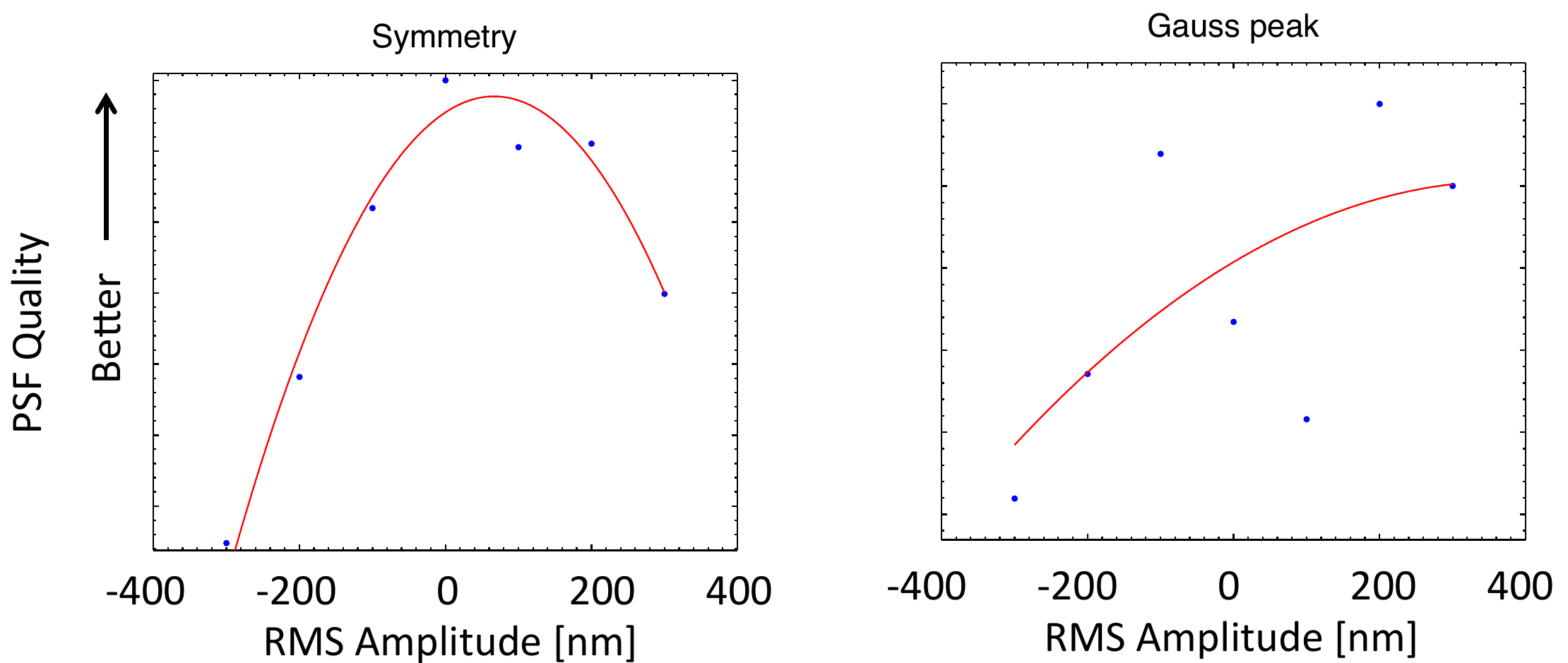}
\end{subfigure}
\newline
\newline
\caption{Comparison of symmetry and peak flux metrics, for a case where the two produce particularly discrepant results. \textbf{Top:} PSF sequence from LMIRCam, both raw frames and PSF azimuthal average-subtracted frames (core masked). The amplitudes of the aberrations span -300~nm RMS to +300~nm RMS.  \textbf{Bottom:} Results of symmetry and peak methods for the sequence above. The peak method mistakenly selects the amplitude corresponding to the image at the far right.}
\label{fig:methodcomp1}
\end{figure}

Although Strehl Ratio is the metric we are ultimately trying to maximize, we find it is not a robust metric for evaluation of PSF quality in short exposures. The ``peak flux'' metric (equivalent to SR) is calculated by fitting a 2D Gaussian profile to the core of the PSF and taking its maximum. The peak flux is quite susceptible to degradation from tip/tilt jitter as well as to brief seeing bursts. Instead, we use a ``symmetry'' metric where we subtract the azimuthal average of the PSF and sum the absolute value of the residuals outside the PSF core.  Figure \ref{fig:methodcomp1} compares the two metrics for one particularly discrepant run, where the symmetry metric returns a sensible value but the peak flux metric does not. The symmetry metric is designed to be sensitive to perturbations of the brightest Airy ring(s). Although it is well-suited to the effects of low-order NCPA, a different metric may be necessary for extension to high-order NCPA, whose effects are evident only in lower signal-to-noise regions of the PSF. We note that azimuthal asymmetry is insensitive to modes such as focus or sphere, except that the effects of other modes may become more pronounced as the PSF is defocused. Therefore, a hybrid approach, using peak flux for focus-like modes and symmetry for all others, is optimal in some cases. 

Several tests of this approach remain. First, we have only had the opportunity to test the eye doctor routine at $4~\mu m$ (both narrowband and broadband filters) and with astigmatism, trefoil, and coma. The addition of more modes is trivial and we expect it to yield further improvements in SR. Second, we wish to determine what improvements can be realized at near-IR wavelengths. A simple scaling with the Marechal approximation tells us that 90\% SR at $4~\mu m$ corresponds to SR of 30--70\% at J--K bands, and this is left to future testing. Third, we have only tested on $R<7$ stars. For efficiency, we would ideally take science camera exposures of 1~sec or less. We expect to have sufficient S/N on the first Airy ring in a single image for $L<8.5-9$ stars. This limit is set by the photon noise from the bright sky background.  We estimate that at $H$-band, where we are read-noise limited for short exposures, we could reach $H=12.5$~mag.  For long wavelength observations of targets $>9$th mag, it would be expedient to calibrate NCPA in the NIR, because we expect the contribution to NCPA from the filters themselves to be small.

We have uncovered limitations to the eye doctor approach. We executed the script once before observations, followed by longer sets of science camera frames to assess PSF quality and stability. Although some sets had excellent stability, others exhibited rapid degradation. In the worst example, the SR degraded by 10\% in less than 10~min after executing the NCPA startup script. In each case where the PSF degraded, the problem was mostly or entirely with returning trefoil. As mentioned previously, trefoil has a narrower range of ``good'' amplitudes, so is more sensitive to drift than astigmatism. 

We have explored several possibilities. The first is variation related to some environmental parameter. However, over timescales of only a few minutes, there is no significant variation in elevation, azimuth, wind, or temperature. The second possibility is that the NCPA vary across the field of view. We see some evidence of this between the top and bottom of the FOV, although it appears to account for less than half of the observed error. The third is related to the way we inject NCPA corrections into the system via reference slope offsets. As discussed previously, the optical gain of a PWFS varies as the AO correction (PSF quality) varies. However, we continue to apply the same amplitude reference slope offset throughout. In other words, the varying optical gain may be changing the effective amplitude of the NCPA we are injecting into the system.  The dataset with 10\% SR degradation was indeed taken under poor and variable seeing, lending support to this hypothesis. However, not every poor seeing dataset exhibits this issue, so further testing and analysis are needed.

\subsection{Future Work} 
\label{sec:Future}

Future improvements will focus on methods to combat the NCPA variability.  These fall broadly into two categories: upgrades to the eye doctor and alternative sensing routines.  We set the constraint that any new scheme must be passive during observing sequences. Although it is possible to stop an observing sequence to recalibrate, this is undesirable not only from the standpoint of efficiency, but because discreet, large changes in the PSF reduce the efficacy of many high-contrast image processing techniques.

First we explore iterations on the current scheme.  We could characterize NCPA variability across the FOV either with an artificial source, by analyzing archival science data, or with new on-sky data.  We could also consider calibrating NCPA off-sky while injecting artificial turbulence onto the mirror to simulate changing seeing conditions. Finally, we could build a library of low-order mode PSFs, to correlate against on-sky frames.  

Ultimately, a more sophisticated solution will likely be necessary. Here, too, several options exist. First, we could implement a real-time optical gain measurement system. A fast ($>0.2$ times the AO loop frequency), small amplitude (of order 5--10~nm) low-order mode such as focus could be added at the mirror level during closed loop. Because in this case the amplitude is known, the WFS response can be physically calibrated. The ability to inject high frequency modes during closed loop already exists for the purposes of on-sky reconstructor generation \cite{Pinna2012}, and could be adapted for our purposes.

A plethora of other NCPA sensing algorithms have been proposed, and could possibly be implemented in our case. Phase diversity \cite{Gonsalves1982}, where NCPA are deduced from images taken in and out of focus, could be used before an observation set begins. However, although a more elegant initial solution than the eye doctor method, it too cannot account for NCPA/PSF variability. Speckle nulling \cite{Malbet1995, Borde2006, Savransky2012a}, where sinusoidal patterns are driven on the mirror to deconstructively interfere with persistent speckles, while a possible solution, is not optimized to act on low-order aberrations. Some form of NCPA phase reconstruction from focal plane images \cite{Codona2013a, Martinache2013}, may be the best option for sensing low-order aberrations. This methodology has the advantage that it can be run continuously from the science frames themselves during an observation set, likely with similar stellar magnitude limits to the eye doctor method.

\section{Summary} 
\label{sec:Summary}

The Large Binocular Telescope Interferometer and its Adaptive Optics system have been optimized for high-contrast observations at $3-5~\mu m$.  The interferometric channel will deliver $10~\mu m$ contrasts of $10^3-10^4$ at 0.1'', while the imaging channel is already delivering $4~\mu m$ contrasts of $10^4 - 10^5$ at 0.3'' -- 0.75''.  The AO system performs well even in poor seeing, with $4~\mu m$ Strehl Ratios of up to 90\% at 1.5''--2'' seeing.

The primary factor affecting the science camera image quality is Non-Common Path Aberration. We have implemented a grid search of low-order aberration modes, yielding improvements of up to 5\% in Strehl Ratio. Time-variable NCPA effects present a challenge, and we are beginning to explore options to mitigate these effects.  In the near-term, real-time optical gain measurement may resolve some of the variability, while in the long term phase reconstruction from science camera images is the most promising technique for our application.



\acknowledgments    
 
We gratefully acknowledge the hard work and dedication of past and present members of the LBTI team including: Tom McMahon, Paul Arbo, Teresa Bippert-Plymate, Elwood Downey, Olivier Durney, Paul Grenz, William Hoffmann, Manny Montoya, Mitch Nash, T.\ J.\ Rodigas, and Elliot Solheid.  We also thank the LBTO staff for their help and support during our commissioning and observing runs.
LBTI is funded by a NASA grant in support of the Exoplanet Exploration Program (NSF 0705296). 
VB was supported by the NSF Graduate Research Fellowship Program (DGE-1143953).
Observations reported here were obtained at the LBT Observatory. The LBT is an international collaboration among institutions in the United States, Italy and Germany. LBT Corporation partners are: The University of Arizona on behalf of the Arizona university system; Istituto Nazionale di Astrofisica, Italy; LBT Beteiligungsgesellschaft, Germany, representing the Max-Planck Society, the Astrophysical Institute Potsdam, and Heidelberg University; The Ohio State University, and The Research Corporation, on behalf of The University of Notre Dame, University of Minnesota and University of Virginia.


\bibliography{library}   

\begin{thebibliography}{10}

\bibitem{Hinz2008}
Hinz, P.~M., Solheid, E., Durney, O., and Hoffmann, W.~F., ``{NIC: LBTI's
  nulling and imaging camera},'' {\em Proceedings of SPIE}~{\bf 7013},  701339
  (2008).

\bibitem{Hinz2008a}
Hinz, P.~M., Bippert-Plymate, T., Breuninger, A., Connors, T., Duffy, B.,
  Esposito, S., Hoffmann, W., Kim, J., Kraus, J., McMahon, T., Montoya, M.,
  Nash, R., Durney, O., Solheid, E., Tozzi, A., and Vaitheeswaran, V.,
  ``{Status of the LBT Interferometer},'' {\em Proceedings of SPIE}~{\bf 7013},
   701328 (July 2008).

\bibitem{Hinz2012}
Hinz, P., Arbo, P., Bailey, V., Connors, T., Durney, O., Esposito, S.,
  Hoffmann, W., Jones, T., Leisenring, J., Montoya, M., Nash, M., McMahon, T.,
  Pinna, E., Puglisi, A., Skemer, A., Skrutskie, M., and Vaitheeswaran, V.,
  ``{First AO-corrected interferometry with LBTI: steps towards routine
  coherent imaging observations},'' {\em Proceedings of SPIE}~{\bf 8445},
  84450U (Sept. 2012).

\bibitem{Hinz2014}
Hinz, P.~M., Bailey, V.~P., Defr\`{e}re, D., Downey, E.~C., Esposito, S., Hill,
  J.~M., Hoffmann, W.~F., Montoya, M., McMahon, T., Puglisi, A.~T., Skemer,
  A.~J., Skrutskie, M.~F., and Vaitheeswaran, V., ``{Commissioning the LBTI for
  use as a nulling interferometer and coherent imager},'' {\em Proceedings of
  SPIE}~{\bf 9146},  in press (2014).

\bibitem{Danchi2014}
Danchi, W.~C., Bailey, V., Bryden, G., Defr\`{e}re, D., Haniff, C.~A., Hinz,
  P.~M., Kennedy, G., Mennesson, B., Millan-Gabet, R., Rieke, G.~H., Roberge,
  A., Serabyn, E., Skemer, A., Stapelfeldt, K.~R., Weinberger, A., and Wyatt,
  M., ``{The LBTI hunt for observable signatures of terrestrial planetary
  systems: a key NASA science program on the road to exoplanet imaging
  missions},'' {\em Proceedings of SPIE}~{\bf 9146},  in press (2014).

\bibitem{Skemer2014}
Skemer, A.~J., Apai, D., Bailey, V.~P., Biller, B.~A., Bonnefoy, M., Brandner,
  W., Buenzli, E., Close, L.~M., Crepp, J., Defr\`{e}re, D., Desidera, S.,
  Eisner, J., Esposito, S., Fortney, J., Henning, T. F.~E., Hinz, P.~M.,
  Hofmann, K.-H., Leisenring, J.~M., Males, J.~R., Millan-Gabet, R., Morzinski,
  K.~M., Pascucci, I., Patience, J., Rieke, G.~H., Schertl, D., Schlieder,
  J.~E., Skrutskie, M.~F., Su, K. Y.~L., Weigelt, G.~P., Woodward, C.~E., and
  Zimmerman, N., ``{High contrast imaging at the LBT: the LEECH exoplanet
  imaging survey},'' {\em Proceedings of SPIE}~{\bf 9148},  in press (2014).

\bibitem{Skrutskie2010}
Skrutskie, M.~F., Jones, T., Hinz, P., Garnavich, P., Wilson, J., Nelson, M.,
  and Solheid, E., ``{The Large Binocular Telescope mid-infrared camera
  (LMIRcam): final design and status},'' {\em Proceedings of SPIE}~{\bf 7735},
  77353H (2010).

\bibitem{Leisenring2012}
Leisenring, J.~M., Skrutskie, M.~F., Hinz, P.~M., Skemer, A., Bailey, V.,
  Eisner, J., Garnavich, P., Hoffmann, W.~F., Jones, T., Kenworthy, M.,
  Kuzmenko, P., Meyer, M., Nelson, M., Rodigas, T.~J., Wilson, J.~C., and
  Vaitheeswaran, V., ``{On-sky operations and performance of LMIRcam at the
  Large Binocular Telescope},'' {\em Proceedings of SPIE}~{\bf 8446},  84464F
  (Sept. 2012).

\bibitem{Hoffmann2014}
Hoffmann, W.~F., Hinz, P.~M., Defrere, D., Leisenring, J.~M., Skemer, A.~J.,
  Arbo, P.~A., Montoya, M., and Mennesson, B., ``{Operation and performance of
  the mid-infrared camera, NOMIC, on the Large Binocular Telescope},'' {\em
  Proceedings of SPIE}~{\bf 9147},  in press (2014).

\bibitem{Defrere2014}
Defr\`{e}re, D., Hinz, P.~M., Downey, E.~C., Hill, J.~M., Mennesson, B.,
  Skemer, A.~J., Vaz, A., Ashby, D.~S., Bailey, V.~P., Zappellini, G.~B.,
  Christou, J.~C., Danchi, W.~C., Grenz, P., Hoffmann, W.~F., Leisenring,
  J.~M., McMahon, T., Millan-Gabet, R., Montoya, M., and Vaitheeswaran, V.,
  ``{Co-phasing the Large Binocular telescope: status and performance of
  LBTI/PHASEcam},'' {\em Proceedings of SPIE}~{\bf 9146},  in press (2014).

\bibitem{Esposito2011}
Esposito, S., Riccardi, A., Pinna, E., Puglisi, A., Quir\'{o}s-Pacheco, F.,
  Arcidiacono, C., Xompero, M., Briguglio, R., Agapito, G., Busoni, L., Fini,
  L., Argomedo, J., Gherardi, A., Brusa, G., Miller, D.~L., Guerra, J.~C.,
  Stefanini, P., and Salinari, P., ``{Large Binocular Telescope Adaptive Optics
  System: new achievements and perspectives in adaptive optics},'' {\em
  Proceedings of SPIE}~{\bf 8149},  814902 (2011).

\bibitem{Riccardi2010}
Riccardi, A., Xompero, M., Briguglio, R., Quir\'{o}s-Pacheco, F., Busoni, L.,
  Fini, L., Puglisi, A., Esposito, S., Arcidiacono, C., Pinna, E., Ranfagni,
  P., Salinari, P., Brusa, G., Demers, R., Biasi, R., and Gallieni, D., ``{The
  adaptive secondary mirror for the Large Binocular Telescope: optical
  acceptance test and preliminary on-sky commissioning results},'' {\em
  Proceedings of SPIE}~{\bf 7736},  77362C (July 2010).

\bibitem{Christou2014}
Christou, J.~C., {Brusa Zappellini}, G., {Guerra Ramon}, J.~C., Miller, D.~L.,
  Wagner, M., and Lefebvre, M.~J., ``{Living with adaptive secondary mirrors:
  365/7/24},'' {\em Proceedings of SPIE}~{\bf 9148},  in press (2014).

\bibitem{Lloyd-Hart2000}
Lloyd-Hart, M., ``{Thermal performance enhancement of adaptive optics by use of
  a deformable secondary mirror},'' {\em Publications of the Astronomical
  Society of the Pacific}~{\bf 112},  264--272 (2000).

\bibitem{Tozzi2008a}
Tozzi, A., Stefanini, P., Pinna, E., and Esposito, S., ``{The double pyramid
  wavefront sensor for LBT},'' {\em Proceedings of SPIE}~{\bf 7015},
  701558--701558--9 (2008).

\bibitem{Esposito2010}
Esposito, S., Riccardi, A., Fini, L., Puglisi, A.~T., Pinna, E., Xompero, M.,
  Briguglio, R., Quir\'{o}s-Pacheco, F., Stefanini, P., Guerra, J.~C., Busoni,
  L., Tozzi, A., Pieralli, F., Agapito, G., Brusa-Zappellini, G., Demers, R.,
  Brynnel, J., Arcidiacono, C., and Salinari, P., ``{First light AO (FLAO)
  system for LBT: final integration, acceptance test in Europe and preliminary
  on-sky commissioning results},'' {\em Proceedings of SPIE}~{\bf 7736},
  773609 (July 2010).

\bibitem{Skemer2012}
Skemer, A.~J., Hinz, P.~M., Esposito, S., Burrows, A., Leisenring, J.,
  Skrutskie, M., Desidera, S., Mesa, D., Arcidiacono, C., Mannucci, F.,
  Rodigas, T.~J., Close, L., McCarthy, D., Kulesa, C., Agapito, G., Apai, D.,
  Argomedo, J., Bailey, V., Boutsia, K., Briguglio, R., Brusa, G., Busoni, L.,
  Claudi, R., Eisner, J., Fini, L., Follette, K.~B., Garnavich, P., Gratton,
  R., Guerra, J.~C., Hill, J.~M., Hoffmann, W.~F., Jones, T., Krejny, M.,
  Males, J., Masciadri, E., Meyer, M.~R., Miller, D.~L., Morzinski, K., Nelson,
  M., Pinna, E., Puglisi, A., Quanz, S.~P., Quiros-Pacheco, F., Riccardi, A.,
  Stefanini, P., Vaitheeswaran, V., Wilson, J.~C., and Xompero, M., ``{First
  Light LBT AO Images of HR 8799 bcde at 1.6 and 3.3um: New Discrepancies
  between Young Planets and Old Brown Dwarfs},'' {\em The Astrophysical
  Journal}~{\bf 753},  14 (2012).

\bibitem{Macintosh2008}
Macintosh, B.~A., Graham, J.~R., Palmer, D.~W., Doyon, R., Dunn, J., Gavel,
  D.~T., Larkin, J., Oppenheimer, B., Saddlemyer, L., Sivaramakrishnan, A.,
  Wallace, J.~K., Bauman, B., Erickson, D.~A., Marois, C., Poyneer, L.~A., and
  Soummer, R., ``{The Gemini Planet Imager: from science to design to
  construction},'' {\em Proceedings of SPIE}~{\bf 7015},  701518 (2008).

\bibitem{Dohlen2006}
Dohlen, K., Beuzit, J.-l., Feldt, M., Mouillet, D., Puget, P., Antichi, J.,
  Baruffolo, A., Baudoz, P., Berton, A., Boccaletti, A., Carbillet, M.,
  Charton, J., Claudi, R., Downing, M., Fabron, C., Feautrier, P., Fedrigo, E.,
  Fusco, T., Gach, J.-l., Gratton, R., Hubin, N., Kasper, M., Langlois, M.,
  Longmore, A., Moutou, C., Petit, C., Pragt, J., Rabou, P., Rousset, G.,
  Saisse, M., Schmid, H.-M., Stadler, E., Stamm, D., Turatto, M., and Wildi,
  F., ``{SPHERE, a Planet Finder instrument for the VLT},'' {\em Proceedings of
  SPIE}~{\bf 6269},  62690Q (2006).

\bibitem{Guyon2010}
Guyon, O., Martinache, F., Garrel, V., Vogt, F., Yokochi, K., and Yoshikawa,
  T., ``{The Subaru Coronagraphic Extreme AO (SCExAO) System: Wavefront Control
  and Detection of Exoplanets with Coherent Light Modulation in the Focal
  Plane},'' {\em Proceedings of SPIE}~{\bf 7736},  773624 (July 2010).

\bibitem{Hinkley2011}
Hinkley, S., Oppenheimer, B.~R., Zimmerman, N., Brenner, D., Parry, I.~R.,
  Crepp, J.~R., Vasisht, G., Ligon, E., King, D., Soummer, R.,
  Sivaramakrishnan, A., Beichman, C., Shao, M., {Roberts, Jr.}, L.~C., Bouchez,
  A., Dekany, R., Pueyo, L., Roberts, J.~E., Lockhart, T., Zhai, C., Shelton,
  C., and Burruss, R., ``{A New High Contrast Imaging Program at Palomar
  Observatory},'' {\em Publications of the Astronomical Society of the
  Pacific}~{\bf 123},  74--86 (2011).

\bibitem{Pinna2012}
Pinna, E., Quir\'{o}s-Pacheco, F., Riccardi, A., Briguglio, R., Puglisi, A.,
  Busoni, L., Arcidiacono, C., Argomedo, J., Xompero, M., Marchetti, E., and
  Esposito, S., ``{First on-sky calibration of an high order adaptive optics
  system},'' {\em Proceedings of SPIE}~{\bf 8447},  84472B (Sept. 2012).

\bibitem{Gonsalves1982}
Gonsalves, R.~A., ``{Phase retrieval and diversity in adaptive optics},'' {\em
  Optical Engineering}~{\bf 21}(5),  829 (1982).

\bibitem{Malbet1995}
Malbet, F., Yu, J.~W., and Shao, M., ``{High-dynamic-range imaging using a
  deformable mirror for space coronography},'' {\em Publications of the
  Astronomical Society of the Pacific}~{\bf 107},  386--398 (1995).

\bibitem{Borde2006}
Bord\'{e}, P. and Traub, W., ``{High-contrast imaging from space: Speckle
  nulling in a low-aberration regime},'' {\em The Astrophysical Journal}~{\bf
  638},  488--498 (2006).

\bibitem{Savransky2012a}
Savransky, D., Macintosh, B.~A., Thomas, S.~J., Poyneer, L.~A., Palmer, D.~W.,
  {De Rosa}, R.~J., and Hartung, M., ``{Focal plane wavefront sensing and
  control for ground-based imaging},'' {\em Proceedings of SPIE}~{\bf 8447},
  84476S (Sept. 2012).

\bibitem{Codona2013a}
Codona, J.~L. and Kenworthy, M., ``{Focal Plane Wavefront Sensing Using
  Residual Adaptive Optics Speckles},'' {\em The Astrophysical Journal}~{\bf
  767},  100 (Apr. 2013).

\bibitem{Martinache2013}
Martinache, F., ``{The Asymmetric Pupil Fourier Wavefront Sensor},'' {\em
  Publications of the Astronomical Society of the Pacific}~{\bf 125},  422--430
  (2013).

\end{thebibliography}
\bibliographystyle{spiebib}   

\end{document}